\documentclass[letterpaper]{article} 
\usepackage{aaai2027} 
\usepackage[hyphens]{url}  
\usepackage{graphicx} 
\usepackage{natbib}  
\usepackage{caption} 
\usepackage{algorithm}
\usepackage{multirow} 
\usepackage{graphicx} 
\usepackage{algorithm} 
\usepackage{algpseudocode}
\usepackage[table]{xcolor} 
\usepackage{colortbl}      
\usepackage{cite}
\usepackage{mathtools}
\usepackage{amsmath,amssymb,bm}
\newcommand{\R}{\mathbb{R}}
\newcommand{\I}{\mathbb{I}}

\newcommand{\cL}{\mathcal{L}}
\newcommand{\vect}[1]{\bm{#1}}

\DeclareMathOperator*{\argmax}{arg\,max}
\usepackage{pifont}

\usepackage{xcolor}

\usepackage{newfloat}
\usepackage{listings}
\DeclareCaptionStyle{ruled}{labelfont=normalfont,labelsep=colon,strut=off} 
\floatstyle{ruled}
\newfloat{listing}{tb}{lst}{}
\floatname{listing}{Listing}

\usepackage{booktabs}

		\title{From Last-Layer to Cross-Layer: One-Shot Joint Classifier Reconstruction for Federated Learning}
		\author{
    Zhenyan Liu\textsuperscript{\rm 1},
    Hua Zhang\textsuperscript{\rm 1},
    Haoran Gao\textsuperscript{\rm 2},\\
    Huiyu Zhou\textsuperscript{\rm 3},
    Zongliang Shen\textsuperscript{\rm 1},
    Qi Li\textsuperscript{\rm 1},\\
    Hongliang Zhu\textsuperscript{\rm 1},
    Yongji Wang\textsuperscript{\rm 1},
    Zi'an Tian\textsuperscript{\rm 1}
}
\affiliations{
    \textsuperscript{\rm 1}Beijing University of Posts and Telecommunications\\
    \textsuperscript{\rm 2}JIUTIAN Research, China Mobile\\
    \textsuperscript{\rm 3}University of Leicester\\
    lzhy@bupt.edu.cn, zhanghua\_288@bupt.edu.cn, gaohaoran@cmjt.chinamobile.com\\
    hz143@leicester.ac.uk, shenzongliang@bupt.edu.cn, liqi2001@bupt.edu.cn\\
    zhuhongliang@bupt.edu.cn, wyj2022@bupt.edu.cn, tza@bupt.edu.cn
}

\begin{document}
			
			\maketitle
			
			\begin{abstract}
				Federated learning (FL) enables collaborative model training while preserving data privacy. However, non-independent and identically distributed (non-IID) data inevitably introduces classifier bias, leading to significant performance degradation. Existing studies primarily focus on calibrating or reconstructing the classifier after FL training to alleviate this bias. We find that the last-layer representations of the global model are often suboptimal for establishing a global decision boundary. Instead, classifiers constructed based on intermediate-layer representations can achieve superior performance. This finding provides a critical insight into overcoming the performance bottleneck inherent in existing paradigms. Nevertheless, identifying the optimal layer without auxiliary datasets or prior knowledge remains a formidable challenge. In this paper, we propose FLARE (\textbf{F}ederated Cross-\textbf{L}ayer Joint Cl\textbf{A}ssifier \textbf{RE}construction) which rethinks classifier reconstruction from a cross-layer perspective and generalizes ridge regression to jointly exploit cross-layer representations, thereby surpassing the performance of any individual layer. Specifically, after FL training, clients perform only a single forward pass to extract class-conditional statistics from predefined layers. The server aggregates these statistics to construct a joint cross-layer ridge classifier, where an adaptive coupling coefficient regulates cross-layer correlations, thereby facilitating the integration of complementary discriminative information. Extensive experiments across 11 datasets, 8 model architectures, and diverse data heterogeneity scenarios demonstrate that FLARE, as a plug-and-play post-processing method, improves accuracy by 11.35\% with only one additional communication round.
				
			\end{abstract}
			

			\section{Introduction}
			
			
			
			

			Federated learning (FL) is a distributed training paradigm that enables collaborative training from multi-source clients without transferring raw data \cite{mcmahan2017communication,li2023edge,zhang2022dense}. However, statistical heterogeneity across clients can skew local decision boundaries toward client-specific class priors, resulting in classifier bias.\cite{luo2021no,zhou2023fedfa,li2023effectiveness}. Aggregating these inconsistent classifiers can substantially degrade the performance of the global  model\cite{fan2023federated,chen2024decoupled,shang2022federated}.
			
			Existing classifier reconstruction approaches in FL can be categorized into two strategies, both of which build upon the global models obtained after FL training or pretrained backbones. The first strategy reconstructs the classifier by leveraging feature statistics uploaded by clients, such as class means, covariance matrices, or regression statistics \cite{goswamicovariances,legate2023guiding,fani2024accelerating,guan2025capture}. The second synthesizes pseudo-features that approximate the global distribution to fine-tune the classifier\begin{figure}[t]
				\centering
				\includegraphics[width=0.95\linewidth]{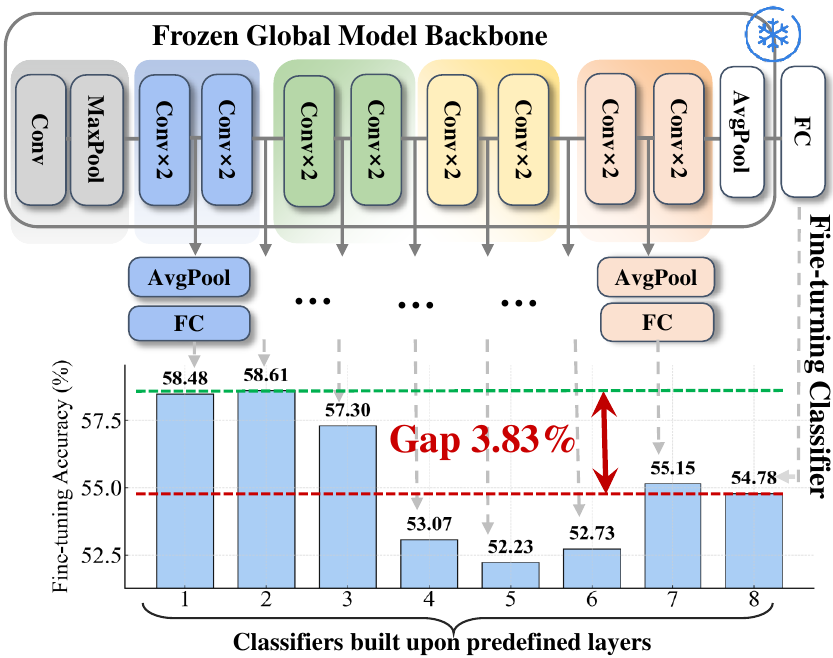}  
				\caption{Top-1 test accuracy of classifiers at various layers. We first train ResNet-18 on CIFAR-10 using FedAvg (100 clients, 5\% participation ratio, Dirichlet distribution with $\alpha=0.1$). Following training, we attach classifiers to predefined layers and fine-tune for 50 epochs on the training set.}
				\label{fig:toy}
			\end{figure} \cite{beitollahifoundation,chen2024decoupled,tang2024dual,pan2024classifier,huang2024federated}. Despite their promise in mitigating classifier bias, these methods are fundamentally limited, as they focus exclusively on last-layer representations. This paradigm implicitly assumes that deeper representations exhibit superior linear separability and semantic discriminability, making them optimal for establishing global decision boundaries\cite{alain2017understanding,cohen2020separability}. However, under non-IID FL conditions, deep features are sensitive to local data distributions, which exacerbates representation divergence across clients and increases instability during aggregation\cite{li2023effectiveness,wang2024fednlr,shi2023understanding}. Consequently, relying solely on last-layer representations undermines the reliability and creates a critical bottleneck for existing reconstruction paradigms.

			To quantify the impact of representation depth on model performance, we conduct a preliminary study. We first train a global model via FedAvg and freeze its backbone. Then, we attach linear classifiers to intermediate layers and fine-tune all classifiers. As illustrated in Figure \ref{fig:toy}, classifiers derived from certain intermediate-layer representations outperform those on the last-layer. This reveals a critical limitation in existing paradigms. For classifier reconstruction, the optimal choice is not strictly limited to the last layer. Instead, it varies across model architectures, datasets, and other factors. However, identifying the optimal layer remains a formidable challenge in FL, where privacy constraints limit access to auxiliary data and prior knowledge\cite{wu2024exploring,zhu2023weak}.

			
			Building upon these findings, we propose FLARE  (\textbf{F}ederated Cross-\textbf{L}ayer Joint Cl\textbf{A}ssifier \textbf{RE}construction) designed to overcome the last-layer decision bottleneck and surpass the optimal-layer performance. FLARE introduces a novel cross-layer 
			perspective for federated classifier reconstruction and further generalizes ridge regression over cross-layer representations, thereby integrating complementary information into a unified decision space. To suppress unreliable cross-layer correlations, FLARE introduces an adaptive coupling coefficient to regulate inter-layer interactions. Consequently, FLARE transcends the limitations of last-layer reconstruction inherent in existing methods, yielding significant improvements in the global model’s performance. Extensive experiments demonstrate that FLARE consistently outperforms state-of-the-art (SOTA) baselines  across various settings. Our contributions are summarized as follows:
			\begin{itemize}
				\item We transcend mainstream FL classifier reconstruction paradigms that focus exclusively on the last layer, revealing the critical value of intermediate-layer representations under data heterogeneity. We accordingly propose FLARE to integrate cross-layer discriminative information for classifier reconstruction. This insight provides a novel perspective on federated classifier reconstruction.

				\item As a plug-and-play method, FLARE further generalizes ridge regression from a single representation space to a jointly optimized cross-layer space. It also introduces an adaptive coupling coefficient to suppress unreliable correlations, thereby constructing a superior joint classifier.

				\item Extensive experiments across five image datasets with three model architectures and six text datasets with five model architectures, covering diverse FL settings, demonstrate that FLARE consistently outperforms existing SOTA methods while preserving privacy.
			\end{itemize}

			\section{Related Work}
			\subsection{FL under Data Heterogeneity}
			Existing research on FL under data heterogeneity follows two primary paradigms: general federated learning (GFL) and Personalized Federated Learning (P-FL). GFL aims to learn a global model applicable to all clients, as exemplified by FedAvg\cite{mcmahan2017communication}. However, non-IID data severely constrains the convergence and performance of the global model. Existing studies have proposed optimization strategies, including mitigating local update drift  \cite{li2020federated,gao2022feddc}, gradient correction\cite{hsu2019measuring,acar2021federated}, and introducing  distillation for knowledge preservation \cite{lu2023federated,lee2022preservation}. In contrast, P-FL seeks to learn client-specific models adapted to each client’s data distributions \cite{smith2017federated}, subsequently inspiring diverse approaches including local fine-tuning of global models \cite{fallah2020personalized}, personalized classifiers \cite{collins2021exploiting}, jointly learned personalized components \cite{t2020personalized,li2021ditto}, and personalized aggregation weights \cite{huang2021personalized,zhang2023fedala}.

			\subsection{Classifier Reconstruction in FL}
			Existing FL classifier reconstruction methods primarily focus on the post-processing phase with a frozen global backbone. Specifically, CCVR\cite{luo2021no}, FedCRGP \cite{pan2024classifier}, and DFL-FS\cite{chen2024decoupled} adopt a generation-based approach, where clients upload class means and covariances, subsequently server generates pseudo-features to reconstruct the classifier. Similarly, FedPFT\cite{beitollahifoundation} utilizes client-uploaded Gaussian mixture models to synthesize features for classifier retraining. FedNCM\cite{legate2023guiding} simplifies this process by requiring only weighted class means to construct a nearest class mean classifier at the server. Alternatively, Fed3R\cite{fani2024accelerating} leverages a recursive regression algorithm, aggregating client-side regression statistics to construct the classifier. To enhance model reliability and robustness, APH\cite{chen2024watch} samples multiple classifiers and combines them after differentiated fine-tuning. Furthermore, FedCGS\cite{guan2025capture} builds a parameter-free Gaussian Naive Bayes classifier based on local feature statistics, while FedCOF\cite{goswamicovariances} exploits properties of the means distribution for unbiased covariance estimation. However, these methods reconstruct classifiers solely based on the last layer, limiting their performance to a single representation space. In contrast, FLARE reconstructs a joint classifier, enabling it to integrate cross-layer discriminative information.

			\begin{figure*}
				\centering
				\includegraphics[width=1.0\textwidth]{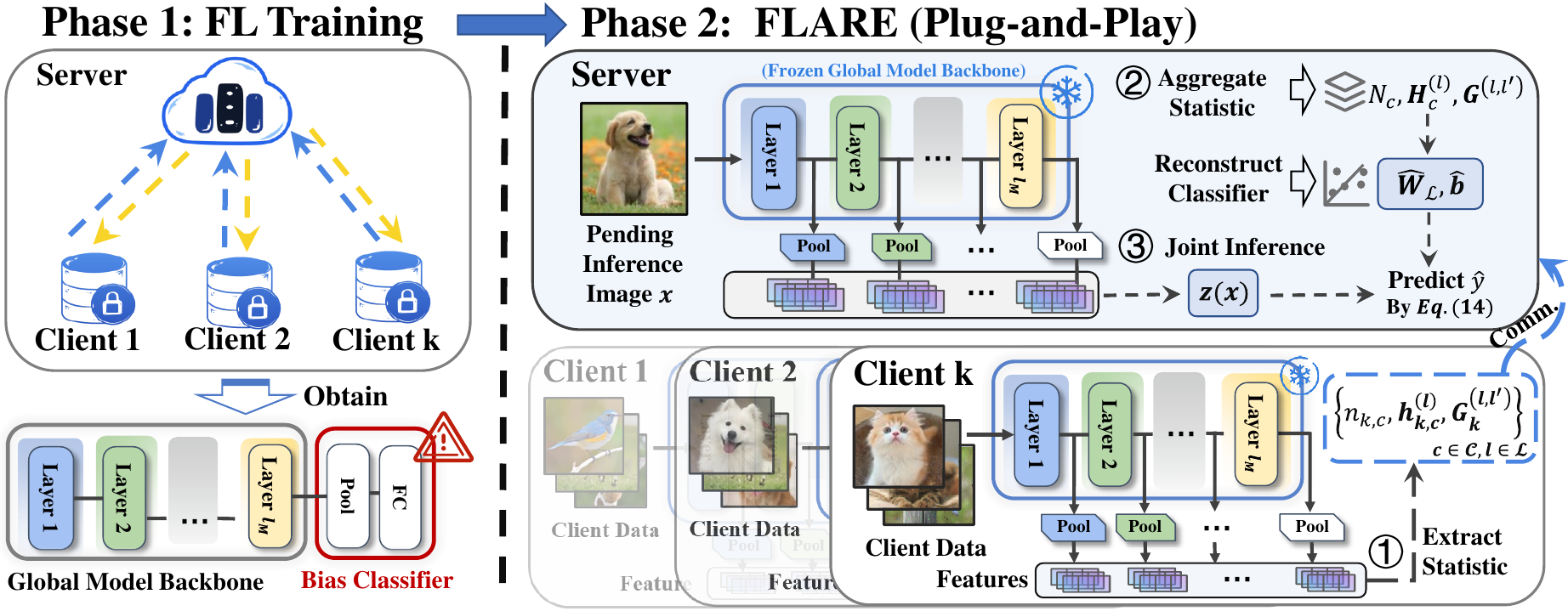}
				\caption{Overview of FLARE. The pipeline consists of two phases: (Phase 1) employs standard FL to obtain a global model, after which the backbone is frozen; (Phase 2) FLARE is deployed as a plug-and-play post-processing method to reconstruct classifier. \textcircled{1} client extracts class-conditional feature statistics from predefined layers and uploads them in one communication round; \textcircled{2} the server aggregates the statistics and constructs cross-layer joint classifier; \textcircled{3} Joint inference with reconstruct classifier}
				\label{fig:framework}
			\end{figure*}

			\section{Method}
			
			\subsection{Problem Formulation}
			
			Consider a FL system comprising $K$ clients indexed by $k \in \{1, \ldots, K\}$. Client $k$ holds a private dataset $\mathcal{D}_k = \{(\mathbf{x}_i^k, y_i^k)\}_{i=1}^{n_k}$, where $y_i^k \in \mathcal{C}$ and $n_k = |\mathcal{D}_k|$. Let $N = \sum_{k=1}^K n_k$ denote the total number of samples across all clients. FL aims to collaboratively learn the parameters $\mathbf{w}$ of a global model without sharing the raw data. The global optimization objective $\mathcal{L}(\mathbf{w})$ is formulated as follows: 
			\begin{equation} \min_{\mathbf{w}} \mathcal{L}_g(\mathbf{w}) = \sum_{k=1}^K \frac{n_k}{N} \mathcal{L}_k(\mathbf{w}) \end{equation} 
			where the local objective of client $k$ is defined as: \begin{equation} \mathcal{L}_k(\mathbf{w}) = \frac{1}{n_k} \sum_{i=1}^{n_k} \ell(f(\mathbf{x}_i^k; \mathbf{w}), y_i^k) \end{equation}
			In practical FL systems, local heterogeneous class priors (i.e., $P_k(y) \neq P(y)$ or $\exists y: P_k(y)=0$ in extreme cases) cause local objectives to systematically deviate from the
			global objective. Under the assumption 
			$\mathcal{P}_k(\mathbf{x}\mid y)=\mathcal{P}(\mathbf{x}\mid y)$, local optimization introduces the gradient discrepancy as :
			\begin{equation}
				\Delta
				=\sum_{c\in\mathcal{C}}\left(P_k(c)-P(c)\right) 
				\cdot\mathbb{E}_{\mathbf{x}\mid c}
				\!\left[\nabla\ell(f(\mathbf{x};\mathbf{w}),c)\right].
			\end{equation}
			This gradient discrepancy can produce a significant disparity in weight magnitudes across classes. After aggregation, the accumulated biases stemming from disparate local distributions distort the learned decision boundaries, ultimately degrading generalization performance.
			
			\subsection{ Overview of FLARE}
			\label{subsec:overview}
			
			To overcome the last-layer reconstruction bottleneck and surpass optimal-layer performance without explicitly identifying it, we develop FLARE, a plug-and-play post-processing method applied after FL training, as illustrated in Figure~\ref{fig:framework}. 
			
			
			\paragraph{One-Shot Extraction of Cross-Layer Statistics.}
			\label{subsec:statistics_collection} Given the model  $f(\cdot; \mathbf{w})$ obtained after FL training, we predefine a set of layers $\mathcal{L} = \{l_1, \dots, l_M\}$. Let $\phi_l:\mathcal{X}\rightarrow\R^{d_l}$ denote the feature mapping for each layer $l\in\cL$, and the activation of input $\vect{x}$ at layer $l_i$ is pooled and flattened into $\vect{z}^{l}=\phi_l(\vect{x})$. Client $k$ computes the class-conditional first moment:
			\begin{equation}
				\label{eq:local_class_count}
				\vect{h}_{k,c}^{(l)} = \sum_{i=1}^{n_k} \I\!\left[ y_i^k = c \right] \vect{z}_i^{k,(l)} \in \R^{d_l}, \quad \forall c \in \mathcal{C}, \; l \in \cL.
			\end{equation}
			To capture the self-correlation $\vect{G}_{k}^{(l,l)}$ of features within layer $l$ and the cross-layer correlation $\vect{G}_{k}^{(l,l')}$ between layers
			$l$ and $l'$, client $k$ computes the pairwise second-order moments:
			\begin{equation}
				\label{eq:local_pairwise_moment}
				\vect{G}_{k}^{(l,l')}
				=
				\sum_{i=1}^{n_k}
				\vect{z}_i^{k,(l)}
				\bigl(\vect{z}_i^{k,(l')}\bigr)^{\top}
				\in\R^{d_l\times d_{l'}},
				\quad l,l'\in\cL.
			\end{equation}
			Client $k$ uploads $
			\mathcal{T}_k=\bigl\{n_{k,c},\,
			\vect{h}_{k,c}^{(l)},\,
			\vect{G}_{k}^{(l,l')}\}_{c\in\mathcal{C},\, l,l' \in \mathcal{L}}
			$ to the server. This procedure requires only one forward pass and one additional communication round.

			\paragraph{Classifier Reconstruction.} The server aggregates the uploaded class-statistics as follows:
			\[
			\label{eq:global_statistics}
			N_c = \sum_{k=1}^{K} n_{k,c},\;
			\vect{H}_{c}^{(l)}=\sum_{k=1}^{K} \vect{h}_{k,c}^{(l)},\;
			\vect{G}^{(l,l')}=\sum_{k=1}^{K} \vect{G}_{k}^{(l,l')}.
			\]
			Let $D=\sum_{l\in\cL}d_l$ and define the global class vector as $\vect{n}=\operatorname{col}_{c\in\mathcal{C}}(N_c)\in\R^{|\mathcal{C}|}$. We further construct the cross-layer first-moment matrix and global feature-sum vector as
			\begin{equation}
				\vect{H}=\operatorname{col}_{l\in\cL}\!
				\left[\vect{H}_{c}^{(l)}\right]_{c\in\mathcal{C}}
				,\;\;
				\vect{H}_{\Sigma}=\operatorname{col}_{l\in\cL}\!
				\sum_{c\in\mathcal{C}}\vect{H}_{c}^{(l)}.
			\end{equation}
			where $\vect{H}\in\R^{D\times |\mathcal{C}|}, \vect{H}_{\Sigma}\in\R^{D}.$Unlike only last-layer reconstruction, $\vect{H}$ retains class-conditional information across all predefined layers. To mitigate the influence of the global feature and label means, we center the statistics as follows:

			\begin{equation}
				\label{eq:centered_joint_statistics}
				\vect{S}^{(l,l')}
				=\vect{G}^{(l,l')}
				-\frac{1}{N}\vect{H}_{\Sigma}^{(l)}
				\bigl(\vect{H}_{\Sigma}^{(l')}\bigr)^{\top}, \;
				\vect{T}
				=\vect{H}-\frac{1}{N}\vect{H}_{\Sigma}\vect{n}^{\top}.
			\end{equation}
			
			\textbf{\textit{(1) Adaptive Coupling:}} To integrate discriminative information while suppressing unreliable inter-layer dependencies under data imbalance, FLARE regulates inter-layer interactions via an adaptive coupling coefficient. The effective global class support is summarized by the harmonic mean
			$
			\label{eq:harmonic_support}
			n_{\mathrm{H}}=|\mathcal{C}|/{\sum_{c\in\mathcal{C}}N_c^{-1}}.
			$
			Compared with the arithmetic mean, $n_{\mathrm{H}}$ is more sensitive to poorly represented classes and therefore better reflects the reliability of cross-layer statistics. The adaptive coupling coefficient is defined as:
			\begin{equation}
				\beta = \frac{\log\left(1 + n_{\mathrm{H}}/|\mathcal{C}|\right)}{\log\left(1 + n_{\mathrm{H}}\right)}  \in[0,1].
			\end{equation}
			The joint second-order matrix is defined by
			\begin{equation}
				\label{eq:adaptive_joint_second-order}
				[\vect{S}_{\beta}]_{l,l'}
				=
				\begin{cases}
					\vect{S}^{(l,l')}, & l=l',\\[2pt]
					\beta\vect{S}^{(l,l')}, & l\neq l',
				\end{cases}
				\qquad l,l'\in\mathcal{L}.
			\end{equation}
			Equivalently, $\vect{S}_{\beta}=(1-\beta)\vect{S}_{\mathrm{diag}}+\beta\vect{S}_{\mathrm{full}}$ preserves  within-layer  second-order, whereas the less reliable 
			inter-layer interactions are regulated. 
			
			\textbf{\textit{(2) Cross-Layer Joint Reconstruction}}: FLARE generalizes ridge regression~\cite{hoerl1970ridge} from a single space to a jointly optimized cross-layer space, where the layer-specific classifiers are coupled through the inter-layer blocks. The resulting cross-layer joint ridge objective is
			\begin{equation}
				\label{eq:adaptive_joint_ridge_objective}
				\begin{aligned}
					\mathcal{J}(\vect{W}_{\mathcal L})
					&=\operatorname{tr}\!\left(
					\vect{W}_{\mathcal L}^{\top}\vect{S}_{\beta}\vect{W}_{\mathcal L}
					\right)
					-2\operatorname{tr}\!\left(\vect{W}_{\mathcal L}^{\top}\vect{T}\right)
					+\lambda\|\vect{W}_{\mathcal L}\|_{F}^{2}.
				\end{aligned}
			\end{equation}
			The off-diagonal blocks of $\vect{S}_{\beta}$ couple the layer-specific classifiers $\vect{W}^{(l)}$  within $ \vect{W}_{\mathcal L}$, revealing the joint nature of the optimization. Thus, rather than being optimized independently or restricted to the last layer, the $\vect{W}^{(l)}$ are coupled via cross-layer terms.The objective first-order optimality condition is 
			\begin{equation}
				\label{eq:joint_ridge_stationarity}
				\nabla\mathcal{J}(\widehat{\vect{W}}_{\mathcal L})
				=2\left(\vect{S}_{\beta}+\lambda\vect{I}_{D}\right)
				\widehat{\vect{W}}_{\mathcal L}-2\vect{T}
				=\vect{0}.
			\end{equation}
			
			Let $\bar{\vect{z}}=\vect{H}_{\Sigma}/N$ and $\bar{\vect{y}}=\vect{n}/N$ denote the global feature and label means, respectively. The intercept is recovered by enforcing zero empirical mean residuals as follows:
			\begin{equation}
				\label{eq:intercept_stationarity}
				\widehat{\vect{W}}_{\mathcal L}^{\top}\bar{\vect{z}}
				+\widehat{\vect{b}}-\bar{\vect{y}}=\vect{0}
				\quad\Longleftrightarrow\quad
				N\widehat{\vect{b}}
				+\widehat{ \vect{W}}_{\mathcal L}^{\top}\vect{H}_{\Sigma}
				-\vect{n}=\vect{0}.
			\end{equation}
			Solving the normal equations in \eqref{eq:joint_ridge_stationarity} and \eqref{eq:intercept_stationarity} yields the unique solution as follows:
			\begin{equation}
				\label{eq:joint_fedridge_solution}
				\left(\vect{S}_{\beta}+\lambda\vect{I}_{D}\right)
				\widehat{ \vect{W}}_{\mathcal L}=\vect{T},
				\quad
				\widehat{\vect{b}}
				=\frac{1}{N}\vect{n}
				-\widehat{\vect{W}}_{\mathcal L}^{\top}\frac{1}{N}\vect{H}_{\Sigma}.
			\end{equation}

			\paragraph{Joint inference.}
			\label{subsec:joint_inference.} For an input $x$, FLARE extracts features $
			\vect{z}(\vect{x})
			=\operatorname{col}_{l\in\mathcal{L}}
			\!\left(\phi_l(\vect{x})\right)\in\R^{D}
			$ from layers in $ \mathcal{L} $. The reconstructed classifier then computes the class score as follows:
			\begin{equation}
				\label{eq:joint_prediction}
				\widehat{y}
				=\argmax_{c\in\mathcal{C}}\widehat{\vect{W}}_{\mathcal L}^{\top}\vect{z}(\vect{x})
				+\widehat{\vect{b}}.
			\end{equation}
			By jointly using cross-layer discriminative information, FLARE is not restricted to the last layer and requires no separate search for the optimal layer. It requires only one statistics-upload round and one server-side linear solve.
			
			\section{Experiment}
			\subsection{Experimental Setup}
			
			\textbf{Datasets and Models.} To comprehensively evaluate the effectiveness of FLARE, we conduct extensive experiments on both image and text classification tasks. \ding{182} \textbf{\textit{Image datasets and models.}} We consider five datasets covering both natural and medical image domains, including SVHN~\cite{netzer2011reading}, CIFAR-10~\cite{krizhevsky2009cifar}, CIFAR-100~\cite{krizhevsky2009cifar}, CINIC-10~\cite{darlow2018cinic}, and BloodMNIST~\cite{yang2023medmnist}. We employ three representative backbone models, namely ResNet-18~\cite{he2016deep}, VGG-11~\cite{simonyan2014very}, and MobileNetV2~\cite{sandler2018mobilenetv2}. \ding{183} \textbf{\textit{Text datasets and models.}} We further evaluate FLARE on six widely used text datasets: AG News~\cite{zhang2015character}, SST-2~\cite{socher2013recursive}, IMDb~\cite{maas2011learning}, 20 Newsgroups~\cite{lang1995newsweeder}, DBpedia-14~\cite{zhang2015character}, and Yahoo Answers Topics~\cite{zhang2015character} on five pre-trained language models, including BERT~\cite{devlin2019bert}, DistilBERT~\cite{sanh2019distilbert}, XLNet~\cite{yang2019xlnet}, ELECTRA~\cite{clark2020electra}, and MiniLM~\cite{wang2020minilm}. For each backbone, we predefined four layers (three intermediate  and the last layer). The layer selection criteria, predefined layer locations, and feature dimensions are detailed in Appendices D and E.

			\textbf{Baselines.} We comprehensively evaluate FLARE as a post-processing method from two perspectives. \ding{182} \textit{\textbf{FL algorithms for the training phase.}} To obtain the global model, we consider the canonical FedAvg and advanced methods designed to mitigate data heterogeneity, including FedDecorr \cite{shi2023understanding}, FedGELA \cite{fan2023federated}, and FedSOL \cite{lee2024fedsol}. \ding{183} \textit{\textbf{Classifier reconstruction methods for post-processing.}} These SOTA baselines serve as direct competitors to FLARE. After FL training with the above algorithms, we freeze the backbone and apply CCVR \cite{luo2021no}, Fed3R \cite{fani2024accelerating}, APH \cite{chen2024watch}, FedPFT  \cite{beitollahifoundation}, FedCOF \cite{goswamicovariances}, and FedCGS \cite{guan2025capture} to reconstruct the classifier. Although several methods assume pre-trained models, we adapt them to the same cold-start setting as our baselines for a fair comparison. Additional comparisons with other methods can be found in Appendix G.

			\begin{table*}[h]
				\centering
				\small
				\caption{Overall comparison of classifier reconstruction methods on ResNet-18 under non-IID settings (default $\alpha=0.1$).}
				\begin{tabular}{l c@{\hspace{5.8pt}} c c c c@{\hspace{5.8pt}} c@{\hspace{5.8pt}} c@{\hspace{5.8pt}} c@{\hspace{5.8pt}} c}
					\noalign{\hrule}
					\multirow{2}{*}{Method} & \multirow{2}{*}{BloodMNIST} & \multicolumn{3}{c}{CIFAR-10} & \multirow{2}{*}{CIFAR-100} & \multirow{2}{*}{CINIC-10} & \multirow{2}{*}{SVHN} & \multirow{2}{*}{Average} \\
					\cline{3-5}
					&  & $\alpha{=}0.05$ & $\alpha{=}0.1$ & $\alpha{=}0.3$ &  &  &  &  \\
					\hline
					FedAvg & 85.09 & 29.83 & 49.64 & 62.74 & 42.55 & 44.14 & 76.76 & -- \\
					\hline
					\hspace{0.8em}+CCVR & 86.42(+1.33) & 38.72(+8.89) & 54.96(+5.32) & 63.09(+0.35) & 42.15(-0.40) & 49.65(+5.51) & 80.87(+4.11) & 3.59 \\
					\hspace{0.8em}+FedPFT & 88.17(+3.08) & 39.43(+9.60) & 54.62(+4.98) & 62.86(+0.12) & 41.23(-1.32) & 47.52(+3.38) & 81.25(+4.48) & 3.48 \\
					\hspace{0.8em}+Fed3R & 89.59(+4.50) & 46.81(+16.98) & 57.40(+7.76) & 64.01(+1.27) & 42.35(-0.20) & 50.71(+6.57) & 83.44(+6.68) & 6.22 \\
					\hspace{0.8em}+APH & 85.44(+0.35) & 31.60(+1.77) & 54.60(+4.96) & 62.97(+0.23) & 42.47(-0.08) & 45.10(+0.96) & 78.73(+1.96) & 1.45 \\
					\hspace{0.8em}+FedCOF & 81.35(-3.74) & 31.34(+1.51) & 49.48(-0.16) & 64.17(+1.43) & 42.65(+0.10) & 45.83(+1.68) & 78.78(+2.02) & 0.41 \\
					\hspace{0.8em}+FedCGS & 90.09(+5.00) & 46.89(+17.06) & 57.47(+7.83) & 63.59(+0.85) & 41.34(-1.21) & 50.39(+6.25) & 83.14(+6.37) & 6.02 \\
					\hspace{0.8em}\textbf{+FLARE} & \cellcolor{gray!10}\textbf{93.66(+8.56)} & \cellcolor{gray!10}\textbf{58.18(+28.35)} & \cellcolor{gray!10}\textbf{64.54(+14.90)} & \cellcolor{gray!10}\textbf{68.46(+5.72)} & \cellcolor{gray!10}\textbf{46.97(+4.42)} & \cellcolor{gray!10}\textbf{56.45(+12.31)} & \cellcolor{gray!10}\textbf{86.19(+9.43)} & \cellcolor{gray!10}\textbf{11.96} \\
					\hline
					FedDecorr & 88.72 & 39.16 & 57.06 & 62.39 & 44.94 & 43.85 & 83.07 & -- \\
					\hline
					\hspace{0.8em}+CCVR & 90.28(+1.56) & 48.95(+9.79) & 62.83(+5.77) & 62.56(+0.17) & 44.43(-0.50) & 48.77(+4.92) & 85.78(+2.71) & 3.49 \\
					\hspace{0.8em}+FedPFT & 90.75(+2.04) & 50.23(+11.07) & 62.66(+5.60) & 62.10(-0.29) & 43.57(-1.37) & 46.71(+2.86) & 86.01(+2.93) & 3.26 \\
					\hspace{0.8em}+Fed3R & 90.82(+2.10) & 53.15(+13.99) & 64.70(+7.64) & 62.94(+0.55) & 43.79(-1.15) & 49.64(+5.79) & 87.13(+4.06) & 4.71 \\
					\hspace{0.8em}+APH & 89.45(+0.73) & 40.87(+1.71) & 61.50(+4.44) & 62.62(+0.23) & 44.88(-0.06) & 44.39(+0.55) & 84.79(+1.71) & 1.33 \\
					\hspace{0.8em}+FedCOF & 87.72(-0.99) & 44.41(+5.25) & 58.46(+1.40) & 63.83(+1.44) & 44.92(-0.02) & 47.21(+3.37) & 85.21(+2.13) & 1.80 \\
					\hspace{0.8em}+FedCGS & 90.41(+1.70) & 53.27(+14.11) & 64.21(+7.16) & 62.98(+0.59) & 42.66(-2.28) & 49.41(+5.57) & 86.95(+3.88) & 4.39 \\
					\hspace{0.8em}\textbf{+FLARE} & \cellcolor{gray!10}\textbf{93.89(+5.17)} & \cellcolor{gray!10}\textbf{60.27(+21.11)} & \cellcolor{gray!10}\textbf{69.46(+12.40)} & \cellcolor{gray!10}\textbf{67.49(+5.10)} & \cellcolor{gray!10}\textbf{47.76(+2.82)} & \cellcolor{gray!10}\textbf{55.08(+11.23)} & \cellcolor{gray!10}\textbf{88.81(+5.74)} & \cellcolor{gray!10}\textbf{9.08} \\
					\hline
					FedGELA & 80.53 & 42.69 & 58.68 & 65.24 & 42.76 & 49.83 & 85.93 & -- \\
					\hline
					\hspace{0.8em}+CCVR & 81.02(+0.49) & 42.93(+0.23) & 58.64(-0.04) & 65.32(+0.08) & 41.98(-0.78) & 50.43(+0.60) & 86.19(+0.26) & 0.12 \\
					\hspace{0.8em}+FedPFT & 83.58(+3.05) & 46.30(+3.61) & 60.98(+2.30) & 65.50(+0.26) & 43.36(+0.60) & 51.90(+2.07) & 87.08(+1.15) & 1.86 \\
					\hspace{0.8em}+Fed3R & 81.47(+0.94) & 43.21(+0.52) & 59.20(+0.52) & 65.12(-0.12) & 43.28(+0.52) & 51.41(+1.58) & 86.58(+0.65) & 0.66 \\
					\hspace{0.8em}+APH & 80.79(+0.25) & 42.51(-0.18) & 58.44(-0.24) & 65.30(+0.06) & 41.22(-1.54) & 50.87(+1.04) & 86.39(+0.46) & -0.02 \\
					\hspace{0.8em}+FedCOF & 79.86(-0.67) & 42.76(+0.06) & 58.53(-0.15) & 65.54(+0.30) & 42.59(-0.17) & 51.14(+1.31) & 86.49(+0.56) & 0.18 \\
					\hspace{0.8em}+FedCGS & 83.48(+2.95) & 46.03(+3.34) & 60.39(+1.71) & 65.34(+0.10) & 44.75(+1.99) & 51.65(+1.82) & 86.68(+0.74) & 1.81 \\
					\hspace{0.8em}\textbf{+FLARE} & \cellcolor{gray!10}\textbf{92.60(+12.07)} & \cellcolor{gray!10}\textbf{62.82(+20.12)} & \cellcolor{gray!10}\textbf{69.65(+10.97)} & \cellcolor{gray!10}\textbf{70.54(+5.30)} & \cellcolor{gray!10}\textbf{47.92(+5.16)} & \cellcolor{gray!10}\textbf{58.74(+8.91)} & \cellcolor{gray!10}\textbf{90.60(+4.67)} & \cellcolor{gray!10}\textbf{9.60} \\
					\hline
					FedSOL & 76.03 & 40.19 & 51.92 & 64.70 & 43.17 & 45.34 & 83.62 & -- \\
					\hline
					\hspace{0.8em}+CCVR & 79.96(+3.93) & 44.12(+3.93) & 53.36(+1.44) & 64.91(+0.21) & 43.44(+0.27) & 48.58(+3.25) & 86.37(+2.75) & 2.25 \\
					\hspace{0.8em}+FedPFT & 82.94(+6.91) & 46.05(+5.86) & 54.12(+2.20) & 65.00(+0.30) & 40.97(-2.20) & 45.42(+0.08) & 86.53(+2.91) & 2.29 \\
					\hspace{0.8em}+Fed3R & 86.64(+10.61) & 51.98(+11.79) & 58.16(+6.24) & 64.81(+0.11) & 43.39(+0.22) & 48.93(+3.60) & 87.42(+3.80) & 5.19 \\
					\hspace{0.8em}+APH & 77.15(+1.12) & 36.68(-3.51) & 52.24(+0.32) & 64.62(-0.08) & 42.69(-0.48) & 46.63(+1.30) & 85.40(+1.78) & 0.06 \\
					\hspace{0.8em}+FedCOF & 76.67(+0.64) & 40.39(+0.20) & 49.54(-2.38) & 64.21(-0.49) & 42.36(-0.81) & 45.84(+0.50) & 85.46(+1.84) & -0.07 \\
					\hspace{0.8em}+FedCGS & 87.39(+11.36) & 52.34(+12.15) & 57.30(+5.38) & 64.75(+0.05) & 42.40(-0.77) & 48.65(+3.31) & 87.55(+3.93) & 5.06 \\
					\hspace{0.8em}\textbf{+FLARE} & \cellcolor{gray!10}\textbf{92.30(+16.27)} & \cellcolor{gray!10}\textbf{62.66(+22.47)} & \cellcolor{gray!10}\textbf{66.22(+14.30)} & \cellcolor{gray!10}\textbf{69.50(+4.80)} & \cellcolor{gray!10}\textbf{46.47(+3.30)} & \cellcolor{gray!10}\textbf{55.47(+10.14)} & \cellcolor{gray!10}\textbf{89.78(+6.16)} & \cellcolor{gray!10}\textbf{11.06} \\
					\hline
					\noalign{\hrule}
				\end{tabular}
				\label{tab:res_res}
			\end{table*}
			\begin{table*}[h]
				\small
				\centering
				\caption{Overall comparison of classifier reconstruction methods on VGG-11 under Dirichlet non-IID settings (default $\alpha=0.1$).}  
				\begin{tabular}{l c@{\hspace{5.5pt}} c c c c@{\hspace{5.5pt}} c@{\hspace{5.5pt}} c@{\hspace{5.5pt}} c@{\hspace{5.5pt}} c}
					\toprule
					\multirow{2}{*}{Method} & \multirow{2}{*}{BloodMNIST} & \multicolumn{3}{c}{CIFAR-10} & \multirow{2}{*}{CIFAR-100} & \multirow{2}{*}{CINIC-10} & \multirow{2}{*}{SVHN} & \multirow{2}{*}{Average} \\
					\cline{3-5}
					&  & $\alpha{=}0.05$ & $\alpha{=}0.1$ & $\alpha{=}0.3$ &  &  &  &  \\
					\hline
					FedAvg & 89.18 & 27.44 & 50.60 & 66.42 & 41.60 & 50.23 & 70.38 & -- \\
					\hline
					\hspace{0.8em}+CCVR & 89.53(+0.34) & 38.27(+10.83) & 54.65(+4.05) & 67.09(+0.67) & 42.40(+0.80) & 53.06(+2.83) & 74.53(+4.15) & +3.38 \\
					\hspace{0.8em}+FedPFT & 89.45(+0.26) & 39.28(+11.84) & 55.11(+4.51) & 67.43(+1.01) & 41.02(-0.58) & 50.94(+0.71) & 72.87(+2.49) & +2.89 \\
					\hspace{0.8em}+Fed3R & 90.83(+1.65) & 44.32(+16.88) & 56.44(+5.84) & 67.35(+0.93) & 37.45(-4.15) & 52.62(+2.40) & 76.02(+5.64) & +4.17 \\
					\hspace{0.8em}+APH & 89.60(+0.42) & 32.80(+5.36) & 53.26(+2.66) & 67.25(+0.83) & 41.53(-0.07) & 51.59(+1.36) & 71.77(+1.38) & +1.71 \\
					\hspace{0.8em}+FedCOF & 87.43(-1.75) & 33.98(+6.54) & 51.36(+0.76) & 66.11(-0.31) & 40.58(-1.02) & 51.59(+1.37) & 72.58(+2.19) & +1.11 \\
					\hspace{0.8em}+FedCGS & 90.59(+1.40) & 44.55(+17.11) & 57.36(+6.76) & 67.66(+1.24) & 40.62(-0.98) & 52.77(+2.54) & 75.83(+5.45) & +4.79 \\
					\hspace{0.8em}\textbf{+FLARE} & \cellcolor{gray!10}\textbf{93.22(+4.03)} & \cellcolor{gray!10}\textbf{62.79(+35.35)} & \cellcolor{gray!10}\textbf{67.04(+16.44)} & \cellcolor{gray!10}\textbf{70.55(+4.13)} & \cellcolor{gray!10}\textbf{44.38(+2.78)} & \cellcolor{gray!10}\textbf{56.79(+6.57)} & \cellcolor{gray!10}\textbf{82.08(+11.70)} & \cellcolor{gray!10}\textbf{+11.57} \\
					\hline
					\hspace{0.8em}FedDecorr & 89.13 & 33.78 & 57.37 & 67.53 & 41.70 & 50.66 & 75.83 & -- \\
					\hline
					\hspace{0.8em}+CCVR & 89.74(+0.61) & 43.01(+9.23) & 60.81(+3.44) & 68.48(+0.95) & 42.01(+0.31) & 53.61(+2.95) & 79.51(+3.68) & +3.03 \\
					\hspace{0.8em}+FedPFT & 88.92(-0.20) & 44.61(+10.83) & 60.07(+2.70) & 68.52(+0.99) & 41.19(-0.51) & 53.52(+2.86) & 80.10(+4.27) & +2.99 \\
					\hspace{0.8em}+Fed3R & 89.40(+0.27) & 44.85(+11.07) & 61.34(+3.97) & 68.34(+0.81) & 33.73(-7.97) & 53.48(+2.82) & 80.39(+4.56) & +2.22 \\
					\hspace{0.8em}+APH & 89.36(+0.23) & 36.94(+3.16) & 59.47(+2.10) & 68.21(+0.68) & 41.91(+0.21) & 51.93(+1.27) & 77.53(+1.70) & +1.34 \\
					\hspace{0.8em}+FedCOF & 87.58(-1.55) & 37.91(+4.13) & 60.25(+2.88) & 68.62(+1.09) & 41.29(-0.41) & 52.05(+1.40) & 78.07(+2.25) & +1.40 \\
					\hspace{0.8em}+FedCGS & 89.92(+0.79) & 46.25(+12.47) & 61.94(+4.57) & 68.75(+1.22) & 40.70(-1.00) & 53.35(+2.69) & 80.03(+4.21) & +3.56 \\
					\hspace{0.8em}\textbf{+FLARE} & \cellcolor{gray!10}\textbf{92.96(+3.83)} & \cellcolor{gray!10}\textbf{62.96(+29.18)} & \cellcolor{gray!10}\textbf{68.67(+11.30)} & \cellcolor{gray!10}\textbf{70.79(+3.26)} & \cellcolor{gray!10}\textbf{44.22(+2.52)} & \cellcolor{gray!10}\textbf{57.87(+7.21)} & \cellcolor{gray!10}\textbf{84.84(+9.01)} & \cellcolor{gray!10}\textbf{+9.47} \\
					\hline
					\hspace{0.8em}FedGELA & 88.51 & 48.86 & 62.05 & 70.09 & 36.88 & 52.51 & 83.35 & -- \\
					\hline
					\hspace{0.8em}+CCVR & 89.59(+1.08) & 56.27(+7.41) & 63.78(+1.73) & 70.44(+0.35) & 41.24(+4.36) & 55.24(+2.73) & 85.68(+2.33) & +2.85 \\
					\hspace{0.8em}+FedPFT & 90.37(+1.86) & 58.47(+9.61) & 63.25(+1.20) & 70.08(-0.01) & 39.82(+2.94) & 51.86(-0.65) & 84.26(+0.91) & +2.27 \\
					\hspace{0.8em}+Fed3R & 89.56(+1.05) & 57.65(+8.79) & 63.40(+1.35) & 68.98(-1.11) & 33.36(-3.52) & 54.79(+2.28) & 84.38(+1.03) & +1.41 \\
					\hspace{0.8em}+APH & 88.11(-0.40) & 41.80(-7.06) & 61.83(-0.22) & 69.63(-0.46) & 36.24(-0.64) & 54.06(+1.55) & 84.79(+1.44) & -0.83 \\
					\hspace{0.8em}+FedCOF & 86.82(-1.70) & 50.26(+1.40) & 59.31(-2.74) & 68.82(-1.27) & 37.17(+0.29) & 54.50(+1.99) & 84.68(+1.33) & -0.10 \\
					\hspace{0.8em}+FedCGS & 90.35(+1.84) & 60.92(+12.06) & 64.20(+2.15) & 70.39(+0.30) & 40.02(+3.14) & 55.05(+2.54) & 86.62(+3.27) & +3.61 \\
					\hspace{0.8em}\textbf{+FLARE} & \cellcolor{gray!10}\textbf{92.96(+4.44)} & \cellcolor{gray!10}\textbf{68.62(+19.76)} & \cellcolor{gray!10}\textbf{70.68(+8.63)} & \cellcolor{gray!10}\textbf{73.21(+3.12)} & \cellcolor{gray!10}\textbf{42.70(+5.82)} & \cellcolor{gray!10}\textbf{58.16(+5.65)} & \cellcolor{gray!10}\textbf{88.36(+5.01)} & \cellcolor{gray!10}\textbf{+7.49} \\
					\hline
					\hspace{0.8em}FedSOL & 84.51 & 33.39 & 57.47 & 68.76 & 36.59 & 50.20 & 76.28 & -- \\
					\hline
					\hspace{0.8em}+CCVR & 86.97(+2.47) & 36.92(+3.53) & 58.89(+1.42) & 68.62(-0.14) & 37.38(+0.79) & 52.38(+2.18) & 79.64(+3.36) & +1.95 \\
					\hspace{0.8em}+FedPFT & 87.57(+3.06) & 40.30(+6.91) & 58.48(+1.01) & 67.81(-0.95) & 36.06(-0.53) & 50.36(+0.17) & 80.52(+4.24) & +1.99 \\
					\hspace{0.8em}+Fed3R & 88.37(+3.86) & 44.44(+11.05) & 58.93(+1.46) & 68.89(+0.13) & 33.37(-3.22) & 52.76(+2.57) & 81.57(+5.29) & +3.02 \\
					\hspace{0.8em}+APH & 86.67(+2.16) & 29.10(-4.29) & 58.83(+1.36) & 68.61(-0.15) & 36.72(+0.13) & 51.18(+0.98) & 78.17(+1.89) & +0.30 \\
					\hspace{0.8em}+FedCOF & 84.98(+0.47) & 33.25(-0.14) & 57.55(+0.08) & 67.95(-0.81) & 35.02(-1.57) & 51.18(+0.98) & 79.35(+3.07) & +0.30 \\
					\hspace{0.8em}+FedCGS & 88.57(+4.06) & 45.31(+11.92) & 59.77(+2.30) & 68.21(-0.55) & 36.36(-0.23) & 52.13(+1.93) & 82.26(+5.98) & +3.63 \\
					\hspace{0.8em}\textbf{+FLARE} & \cellcolor{gray!10}\textbf{92.29(+7.79)} & \cellcolor{gray!10}\textbf{63.31(+29.92)} & \cellcolor{gray!10}\textbf{67.48(+10.01)} & \cellcolor{gray!10}\textbf{71.39(+2.63)} & \cellcolor{gray!10}\textbf{40.92(+4.33)} & \cellcolor{gray!10}\textbf{57.35(+7.16)} & \cellcolor{gray!10}\textbf{84.82(+8.55)} & \cellcolor{gray!10}\textbf{+10.05} \\
					\bottomrule
				\end{tabular}
				\label{tab:vgg_res}
			\end{table*}

			\textbf{Implementation Details.} Following standard FL settings \cite{lu2023federated,lee2022preservation}, we consider two challenging data heterogeneity scenarios: (1) Dirichlet non-IID, where the label distribution is controlled by a parameter $\alpha$ (a smaller $\alpha$ indicates higher skewness); and (2) Pathological non-IID, where each client possesses only $S$ classes. During the FL training phase, all 100 clients employ SGD with momentum of 0.9 and weight decay of $10^{-5}$. The batch size is set to 128, and the initial learning rate is 0.01, which decays exponentially by a factor of 0.99 after each communication round. In each round, 5\% of clients participate in local updates for 5 epochs. To ensure sufficient convergence, the total rounds are set to 500. After training, all classifier reconstruction methods are executed based on the frozen global model backbone. $\lambda$ is set to $10^{-3}$ according to the sensitivity analysis provided in Appendix H. All experiments are implemented in PyTorch (v2.2.1, CUDA 12.1) and run on an Intel Xeon Platinum 8375C CPU and NVIDIA RTX 3090 GPU.

			\subsection{Experimental Results}
			\textbf{Overall Comparison.} Tables~\ref{tab:res_res}--\ref{tab:vgg_res} and Appendix Table 1 compare the accuracy of FLARE against existing SOTA classifier reconstruction methods across three architectures, five datasets, and Dirichlet non-IID settings $\alpha \in \{0.05, 0.1, 0.3\}$. Methods marked with "+" indicate that the reconstruction methods after FL training (e.g., FedAvg), and the reported accuracy is evaluated after this reconstruction. All results are averaged over three random seeds with small standard deviations detailed in Appendix I. The best results are highlighted in bold with a gray background. FLARE consistently and significantly improves the global model performance across all evaluated scenarios, achieving SOTA and competitive results. On average, FLARE yields a $10.18\%$ accuracy improvement over the original global model. Compared with the strongest reconstruction baselines, FLARE further boosts accuracy by $5.03\%$, $4.06\%$, and $8.04\%$ on ResNet-18, VGG-11, and MobileNetV2, respectively. Notably, under extreme heterogeneity ($\alpha=0.05$), FLARE achieves an average accuracy gain of 23.43\% , highlighting its effectiveness in mitigating severe classifier bias. We also evaluate FLARE under the Pathological non-IID setting, with detailed results provided in Appendix B. By exploiting cross-layer discriminative information, FLARE overcomes the performance bottleneck of relying solely on the last layer for reconstruction, thereby achieving consistent and substantial performance gains.

\begin{figure}[t]
				\centering
				\includegraphics[width=0.95\linewidth]{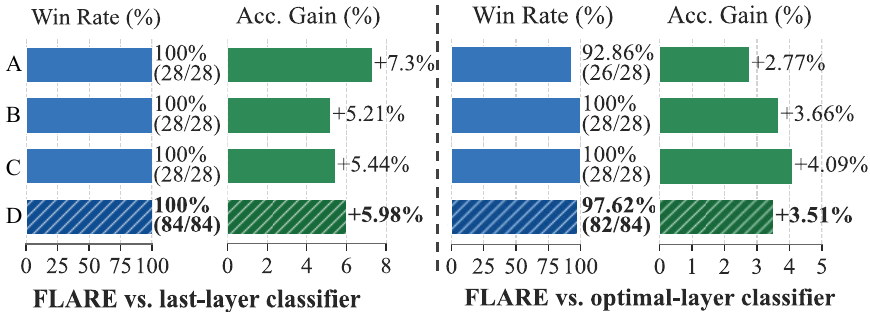}  
				\caption{Win rate and average accuracy gains of FLARE across different architecture tasks. A, B, C, D denote ResNet-18, VGG-11, MobileNetV2 and Overall results, respectively. }
				\label{fig:vs}
			\end{figure}

			\textbf{Advantage of the Cross-layer Strategy.}  We summarize the results of 84 FL tasks in Tables~\ref{tab:res_res}--\ref{tab:vgg_res} and Appendix Table 1. As shown in Figure~\ref{fig:vs}, we compare FLARE with two baselines: (1) last-layer reconstruction (existing SOTA methods) and (2) single optimal-layer reconstruction where the optimal layer is determined by validating each layer individually on the dataset. We report the win rate and the average relative accuracy gain of FLARE across tasks. FLARE surpasses last-layer reconstruction in all tasks, with an average gain of 5.98\%. More notably, FLARE surpasses even  optimal single-layer reconstruction, with an average gain of 3.51\%, without requiring any explicit search for the optimal layer.

			
			\textbf{Comparison on Text Classification Tasks.} As shown in Table \ref{tab:main_text_results}, FLARE consistently outperforms all competing classifier reconstruction methods across diverse datasets and mainstream language model architectures. Compared with the strongest baselines, FLARE achieves accuracy improvements of up to 5.0\% and 14.8\% on BERT and XLNet, respectively, demonstrating its effectiveness across heterogeneous text classification scenarios. Additional results on SST-2, IMDb, and DBpedia-14 with DistilBERT, ELECTRA, and MiniLM architectures have been provided in Appendix C.

			\begin{figure*}[t]
				\includegraphics[width=\textwidth]{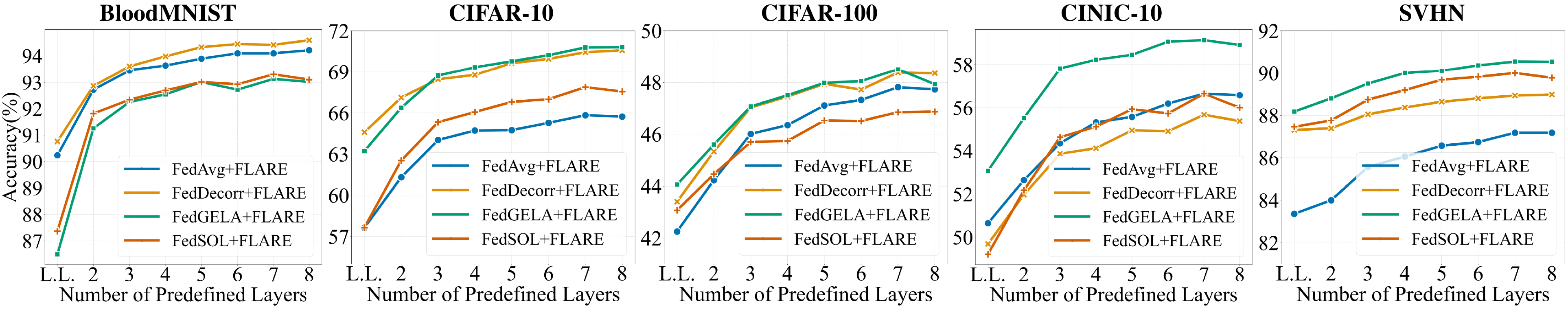}
				\caption{Ablation on the number of predefined layers in FLARE. \textbf{L.L.} indicates the reconstruction only based on the last-layer .}
				\label{fig:layers_num}
			\end{figure*}

			\begin{table}[t]
				\centering
				\renewcommand{\arraystretch}{0.75} 
				\setlength{\tabcolsep}{2.8pt} 
				\small
				\caption{Accuracy comparison on AG News, Yahoo Answers Topics (Yahoo), and 20Newsgroup (20NG) }
				\label{tab:main_text_results}
				\begin{tabular}{l c c c c c c}
					\toprule
					Model & \multicolumn{3}{c}{BERT} & \multicolumn{3}{c}{XLNET} \\
					\cmidrule(lr){2-4} \cmidrule(lr){5-7}
					Dataset & AG News & Yahoo & 20NG & AG News & 20NG & Yahoo \\
					\midrule
					CCVR & 85.70 & 63.08 & 57.82 & 25.00 & 39.58 & 40.27 \\
					FedPFT & 85.08 & 60.40 & 56.45 & 74.84 & 36.18 & 35.50 \\
					Fed3R & 89.63 & 68.59 & 62.47 & 84.87 & 45.65 & 50.46 \\
					APH & 83.84 & 66.05 & 34.35 & 74.96 & 10.57 & 48.29 \\
					FedCOF & 83.33 & 60.54 & 57.21 & 74.88 & 25.80 & 28.47 \\
					FedCGS & 89.70 & 67.80 & 61.46 & 84.76 & 41.73 & 50.36 \\
					\textbf{FLARE} & \cellcolor{gray!10}\textbf{91.36} & \cellcolor{gray!10}\textbf{70.79} & \cellcolor{gray!10}\textbf{66.46} & \cellcolor{gray!10}\textbf{89.74} & \cellcolor{gray!10}\textbf{51.37}& \cellcolor{gray!10}\textbf{65.21} \\
					\bottomrule
				\end{tabular}
			\end{table}
			
			\begin{table}[t]
				\centering
				\renewcommand{\arraystretch}{0.9} 
				\setlength{\tabcolsep}{2.6pt} 
				\small
				\caption{Ablation study on fixed values and our proposed $\beta$.}
				\label{tab:beta_ab}
				\begin{tabular}{l c c c c c c c}
					\toprule
					Dataset & 0.0 & 0.2 & 0.4 & 0.6 & 0.8 & 1.0 & Ours $\beta$ \\
					\midrule
					BloodMNIST & 91.11 & 91.47 & 91.87 & 92.65 & 92.28 & 92.26 & \cellcolor{gray!10}\textbf{92.90} \\
					CIFAR-10 & 61.30 & 62.44 & 62.83 & 62.92 & \cellcolor{gray!10}\textbf{63.19} & 63.02 & 63.13 \\
					CIFAR-100 & 42.35 & 43.31 & 43.08 & 43.05 & 42.61 & 42.08 & \cellcolor{gray!10}\textbf{43.50} \\
					SVHN & 83.42 & 84.31 & 84.62 & 84.21 & 84.11 & 84.02 & \cellcolor{gray!10}\textbf{84.78} \\
					Average & 69.55 & 70.38 & 70.60 & 70.71 & 70.55 & 70.35 & \cellcolor{gray!10}\textbf{71.08} \\
					\bottomrule
				\end{tabular}
			\end{table}
			
			\subsection {Ablation Studies} 
			We investigate the effects of the number of layers and the hyperparameter $\beta$ on FLARE. Figure \ref{fig:layers_num} shows a clear trend: model accuracy improves significantly as the number of predefined layers initially increases. Notably, even incorporating just one additional intermediate layer yields a substantial performance boost over the baseline only based on the last layer. When the number of layers reaches five, the marginal benefits diminish, and performance begins to plateau. This suggests that a moderate number of layers is sufficient to exploit the cross-layer complementary information for stable performance improvement. 
			Table~\ref{tab:beta_ab} evaluates our adaptive $\beta$ against fixed baselines (averaged across three architectures). While the optimal fixed value varies by dataset, $\beta$ yields the best performance on three datasets and remains within 0.06\% of the optimum accuracy. Overall, it achieves the best average accuracy and outperforms the uncoupled baseline by 1.53\%, thereby validating the efficacy of joint cross-layer coupling.

			\begin{figure}[t]
				\centering
				\includegraphics[width=1.0\linewidth]{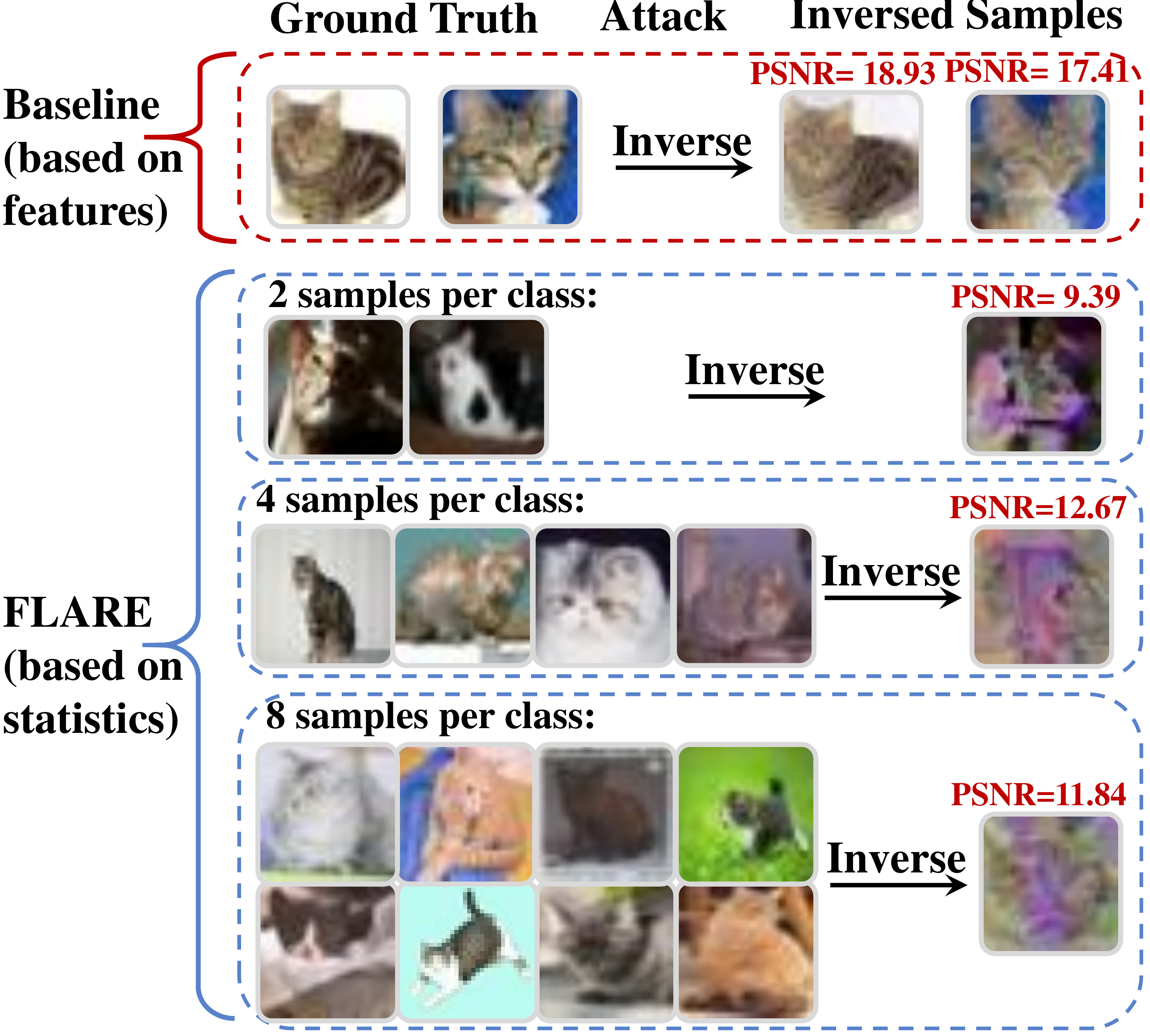}  
				\caption{Comparison of inversion results when the server accesses per-sample raw features vs. the class-conditional statistics uploaded by FLARE (2,4, and 8 samples per class).}
				\label{fig:inverse}
			\end{figure}
			
			\subsection{Analysis}
			\textbf{Privacy Analysis.} Uploading client-side statistics in FLARE may raise privacy concerns. However, FLARE relies only on aggregated class-conditional statistics rather than client specific instance-level information. Consequently, FLARE is naturally compatible with secure aggregation protocols \cite{bonawitz2017practical,mai2024rflpa}, under which the server can only access the aggregated results. This design fundamentally prevents the server from isolating any single client’s contribution and recovering private data from individual uploads. Furthermore, we consider a more challenging scenario without secure aggregation. The server merely observes class-conditional statistics. To empirically validate the privacy guarantee, we evaluate FLARE against a feature inversion attack \cite{mahendran2015understanding}. We simulate an adversarial setting that is highly favorable to the attacker, where the server is assumed to be the adversary and a client holds very few single-class samples, representing an extreme scenario. As shown in Figure \ref{fig:inverse}, when the server has direct access to the raw features of a single sample, the corresponding data can be reconstructed accurately. In contrast, when the attacker is restricted to the statistics uploaded by FLARE, the reconstructed images collapse to unrecognizable noise, even when the client contains only 2,4, and 8 samples for the same class. The consistently low PSNR values provide quantitative evidence of reconstruction failure.

			\section{Conclusion}
			This paper conducts research on the problem of classifier bias in non-IID FL. We present a key observation that the last-layer representations are not always the most suitable for constructing global decision boundaries; rather, certain intermediate representations exhibit stronger generalization capabilities. This finding reveals the inherent performance bottleneck in the prevailing paradigm, which reconstructs classifiers based solely on last-layer representations, and points toward a new direction for classifier reconstruction in FL. To address the non-trivial challenge of precisely locating the optimal intermediate layer under privacy constraints, we propose FLARE, which leverages the cross-layer discriminative information via generalized ridge regression and inter-layer interaction regulation, thereby achieving stable gains while bypassing the difficulty of optimal-layer identification. Experimental results demonstrate that FLARE consistently improves performance across multiple benchmark datasets, model architectures, and various heterogeneous settings.

			\bibliography{aaai2027}

		\end{document}